# From quantitative modeling of fluorescence experiments on biomolecules to the prediction of spectroscopic dye properties

Thomas-Otavio Peulen*, Daria Maksutova and Thorben Cordes*

Biophysical Chemistry, Department of Chemistry and Chemical Biology, Technische Universität Dortmund, Otto-Hahn-Str. 4a, 44227 Dortmund, Germany

*corresponding authors: thomas.peulen@tu-dortmund.de, thorben.cordes@tu-dortmund.de

## Abstract

Fluorescence spectroscopy and modeling provide powerful means to characterize biomacromolecular structures, dynamics, and interactions. Förster resonance energy transfer serves as a key technique for this due to its nanometer-scale distance sensitivity. Quantitative interpretation of fluorescence data relies on models that link molecular structure to observable spectroscopic quantities and vice versa. Integrative modelling frameworks combine fluorescence observables with complementary structural information to infer molecular structures and conformational ensembles. This review outlines conceptual components of fluorescence-based modeling, discusses dye representations, and highlights advances toward refined models enabling quantitative structural analysis. Finally, we discuss the prediction of spectroscopic properties of dyes based on biomolecular structures and fluorescence assay design beyond traditional FRET applications.

## Introduction

Fluorescence experiments and modeling of fluorescence experiments are frequently used to characterize structures, dynamics and interactions of biomacromolecules [1]. Owing to its distance dependence on the nanometer scale, Förster resonance energy transfer (FRET) is well-suited for this (**Fig. 1a**). Since its first quantitative description, structural models have been used to interpret spectroscopic FRET data, initially to describe the spatial chlorophyl A arrangement in photosynthetic systems [2]. Numerous subsequent works extended these concepts. Nowadays, FRET is routinely using in structural modeling complementing classic structural information to infer biomolecular conformations and assemblies [3]. FRET encodes information on static distances and dynamic processes, such as rotational diffusion and conformational exchange in fluorescence decays, correlation spectroscopy, single-molecule histograms, and immobilized single-molecule trajectories (**Fig. 1a**). Quantitatively, all experiment types are interpreted through models.

Modeling generally refers to the construction of mathematical or computational representations that relate physical or structural parameters to experimental observables (**Fig. 1b**). Formally, the space of possible models is sampled and models are ranked based on their agreement with the used information [5]. A classic example is the structure of double stranded DNA that was based on chemical information in combination with X-ray scattering [4].

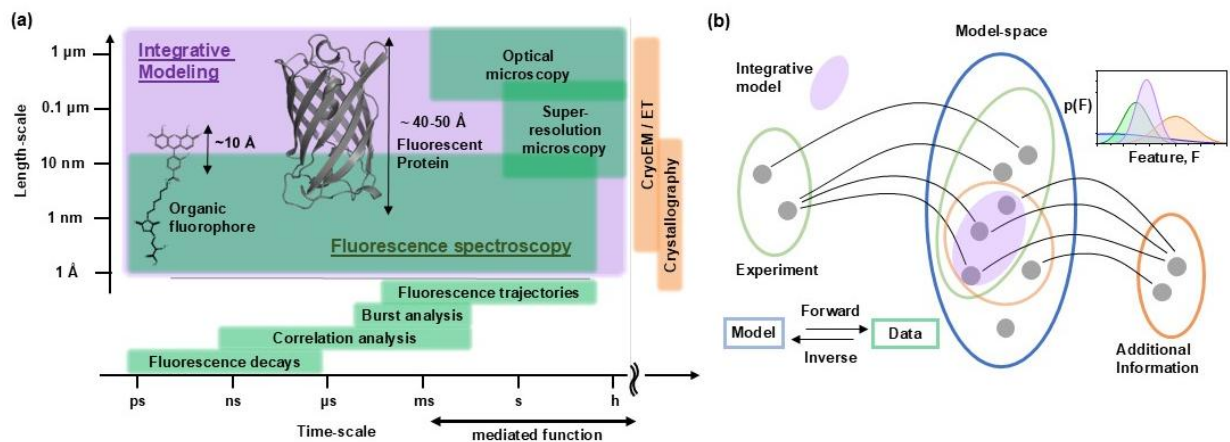

**Figure 1 | Fluorescence in integrative structural modeling including modelling principles**. a) Fluorescence experiments probe biomolecular structure and dynamics from length scales ranging from $10^{-10}$ to $10^{-6}$ m on timescales from picoseconds to hours. Dye representations define the mapping between structural ensembles and experimental observables and ultimately determine the achievable structural resolution across biologically relevant length and time scales. b) Conceptual view of an integrative modeling workflow. In the forward model, candidate structural or kinetic representations are transformed into predicted fluorescence observables. The inverse model identifies structural ensembles that satisfy the experimental restraints within their uncertainties rather than selecting a single optimal solution. Distinct experiments impose complementary restraints on a shared model space; their overlap defines an

integrative ensemble consistent with all available data.

Deviations between model predictions and data arise from experimental uncertainties and limitations of the model representation (model accuracy). Forward models map the conformational representation of a structure or structural model to experimentally observable quantities (**Fig. 1b**). The original treatment by Theodor Förster illustrates this approach [2]. Förster calculated steady-state FRET efficiencies for different structural arrangements under defined assumptions, including rapid rotational diffusion of fluorophores, isotropic orientational averaging, and specific arrangement of chromophore in different lattice geometries. These assumptions define a forward model relating structures to FRET efficiencies, enabling evaluations of structural hypotheses against experiments.

Fluorescence experiments inform on intensities, anisotropies, lifetimes, quantum yields or kinetic parameters of purified biomacromolecules or in live cell experiments down to the single molecule level. Fluorescence intensities allow correlation analysis (fluorescence correlation spectroscopy, FCS), give access to single-molecule histograms or intensity traces of individual molecules to monitor FRET or protein-induced fluorescence enhancement (PIFE) [6–8]. Fluorescence lifetimes and anisotropies can further report on the local dye environment. As such fluorescence investigations enable the study of many biological processes[1–3], including identification of protein conformations [9], resolution of conformational microstates [10] using Markov models on photon traces [11], the characterization of transcription factors [12] and kinetic elements in proteins [13]. Fluorescence experiments integrate well with traditional techniques such cross-linking mass spectrometry [14], small-angle X-ray scattering (SAXS) [15], electron paramagnetic resonance (EPR) [16], or nuclear magnetic resonance (NMR)[17]. Furthermore they can resolve dynamic states in proteins [16], heterogeneous conformational ensembles [18,19] of unfolded proteins [20], and disordered structures like PSD-95 [21,22], lipase foldase [23] alpha synuclein [24], and explain strong interactions in intrinsically disorder proteins [25]. The wide applicability and proven accuracy of fluorescence-based techniques for studying nucleic acids [26] and proteins [27] qualifies them as ideal tool for dynamic studies of biomacromolecules and its complexes *in vitro* but also in living cells.

Quantitative fluorescence experiments integrate multiple representation layers. In single-molecule FRET, experiments registered photons are grouped, FRET efficiencies are computed, and distances are derived. In structural interpretations, forward models, which describes the spatial and dynamic behavior of the attached fluorophores for a structure or ensemble of structures, are a central component. Depending on the level of approximation, dye models compute static spatial distributions, time-dependent conformational sampling, and/or orientational and distance-dependent coupling between dyes. The resulting dye-derived features, such as distance distributions or effective FRET efficiencies, are then compared to experimental observables within the noise model of the experiment (**Fig. 1b**).

In this review, we focus on the modeling that links structural hypotheses to basic properties such as spatial dye distributions and orientations that can be converted to spectroscopic data. This is not only relevant for the understanding of a specific biomolecular system but also for the systematic design of fluorescent assays [28,29]. To describe the modelling, we define conceptual components required for a forward model, emphasizing the conversion from molecular structure to fluorescence observables rather than the inverse transformation. Next, we discuss the molecular dye representations spanning single static conformations to heterogeneous ensembles, together with their associated levels of coarse graining. We then review established dye models highlighting their assumptions, domains of applicability, and limitations. Finally, we outline emerging directions to use structural information for forward modelling beyond FRET in which structural data is used to predict spectroscopic properties of dyes. Such modalities are expected to drive refined dye representations enabling assays considering physico-chemical properties of the dye but also its interactions with (*i*) the solvent, (*ii*) other dyes and (*iii*) the biomolecules to systematically establish fluorescence assays and biosensors.

## Integrative modelling using FRET

Modeling provides a formal representation of a physical system using equations, algorithms, and data structures for quantitative interpretation of experimental observations (**Fig. 1**). In fluorescence spectroscopy, modeling involves several interconnected components: a structural or kinetic model describing the biomolecular system, a forward model that predicts experimental observables for structural representations, and a noise model that accounts for uncertainties arising from both measurement error and model approximations. Often, the underlying structural–kinetic model space is only partly restrained by the experimental and auxiliary information, leading to ambiguity in the inferred representation. Rather than mapping this model space directly, measurements are interpreted through photophysical models to obtain intermediate quantities, such as FRET efficiencies, distance distributions, rate constants, or brightness, that can subsequently be related to structural, ensemble, or kinetic representations. This layered architecture places fluorophore modeling at the center of quantitative interpretation, because structural hypotheses require an accurate description of the dye behavior. For a given molecular representation[1], the forward problem is well defined: fluorescence observables can be computed from the model. The inverse direction, however, is fundamentally ill-posed. Distinct structural or kinetic models may produce indistinguishable fluorescence signatures, rendering the mapping from observables to structure non-bijective. Consequently, the objective of fluorescence modeling is not the identification of a single "correct" structure but the determination of the set or distribution of models consistent with the data (**Fig. 1**). Sampling the space of admissible structures therefore provides the natural solution to the inverse problem, particularly within probabilistic or Bayesian frameworks.

Because fluorescence-based structural modeling constitutes an ill-posed inverse problem, the experimental data alone rarely determine unique structures. Instead, model building typically follows an iterative workflow in which experimental design, data acquisition, and modeling inform each other [30] (**Fig. 2**). The selected examples follow four conceptual stages: (*i*) gathering data, (*ii*) model representation, (*iii*) sampling, and (*iv*) analysis with validation. Experimental information such as sequence connectivity, secondary structure, prior structural knowledge, and fluorescence observables is first assembled to define the modeling problem. The structural representation is then adapted to the system under study; for example, a minimal rigid body decomposition of the maltose binding protein (MalE) is constructed to represent the labeling network defined by the FRET pairs labelled with donor and acceptor dyes at positions 29/352, 36/352, 84/352, 87/186, 134/186, and 34/205 (**Fig. 2a**) [27]. Such decompositions are typically guided by existing structural data and secondary structure, whereas atomistic force fields serve as knowledge in more flexible structures. The chosen representation defines the accessible model space. FRET restraints are incorporated to bias sampling toward conformations that agree with the measurements, typically using Monte Carlo or molecular dynamics strategies to identify well scoring models. During analysis, models are evaluated and posterior distributions are characterized to quantify structural uncertainty. Validation relies on independent information not used during modeling, including orthogonal experiments and general biophysical plausibility. The MalE

---

[1] Note, "structure" in this context may refer either to a single conformation or to a physical ensemble populated in solution. Such ensembles are not artifacts of experimental noise but reflect genuine thermodynamic heterogeneity.

representations are schematic and shown for conceptual purposes only.

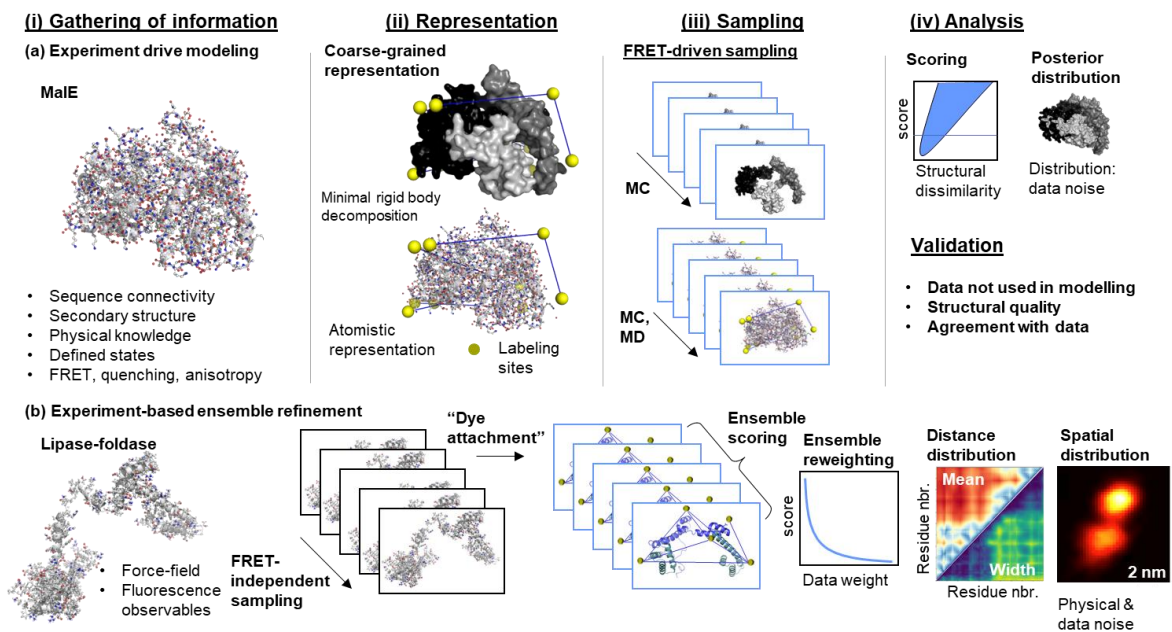

**Figure 2 | Examples for incorporating FRET restraints into structural modeling.** (a) FRET driven sampling of compatible structures. (b) FRET ensemble refinement via maximum entropy. This workflow is conceptually based on the ensemble refinement strategy [23].

In a second strategy (**Fig. 2b**), FRET data are not used during sampling. Instead, a conformational ensemble is generated from a physical prior, commonly a force field, which defines the model space and represents the underlying physical ensemble. After dye attachment and forward modeling, the population of conformers itself becomes the model representation. The entire ensemble is then refined by reweighting using a maximum entropy or maximum a posteriori framework that introduces the minimal perturbation required to achieve agreement with the experimental restraints. Because sampling is decoupled from the forward model, this approach is generally easier to implement and less dependent on computational efficiency. Analysis yields ensemble level properties such as intra residue distance distributions and overall spatial heterogeneity, with the final ensemble determined jointly by experimental noise and the diversity of the sampled physical ensemble. Again, validation relies on independent data and established biophysical knowledge.

The limitations of fluorescence-based structural inference can be summarized in a triangle of frustration that captures the trade-off between (i) the number of independent observations restraining a model, (ii) the level of molecular detail in the representation, and (iii) the effective noise of the model. For a fixed number of free parameters, a sparse spectroscopic data set cannot uniquely resolve atomistic detail. Integrative strategies therefore balance model resolution against data content and uncertainty, consistent with optimal-representation principles developed in integrative structural biology [31]. Experimental success further depends strongly on sample preparation, making deliberate construct and labeling strategies essential. Tools such as Labelizer increase the probability of informative measurements by guiding labeling site selection using evolutionary conservation and experimental knowledge for an improved design [28,29]. At the same time, fluorescence observables report not only on structure but also on conformational dynamics, intermolecular interactions, and dye photophysics; experiments should therefore be engineered to impose maximally discriminative constraints on candidate models. Complementarily, optimized FRET networks can maximize structural information content and improve the resolvability of molecular architectures [32]. Together, rational labeling strategies and information-optimal FRET geometries form a unified experimental design framework that naturally integrates with modeling approaches, where priors, forward models, and experimental configurations are progressively refined.

## Dye models

In practice, fluorescence measurements are rarely linked directly to structural models. Instead, models convert photon-level data into parameters that can be interpreted within structural or kinetic frameworks. Early modeling approaches represented the spatial dye distribution by mean positions to apply restraints to determine structures [33]. However, to fluorescence observables to structures at high accuracy demands an explicit description of dye positions, orientations, and environmental interactions in a dye-model with physical realism (**Fig. 3**). Because fluorescence reports on the dye rather than the biomolecule itself, structural interpretation depends critically on the dye representation. The dye model thus serves as the central bridge between the molecular structure and an observable signal. Atomistic molecular dynamics (MD) simulations with parameterized dyes [34,35] can be regarded as the most detailed and accurate dye representation [36]. By simulating continuous trajectories, MD explicitly captures many relevant dye interactions and, in principle, resolves a large part of positional, orientational, and photophysical heterogeneity. In practice, however, the substantial computational cost restricts its use in large-scale sampling, uncertainty quantification, and integrative modeling workflows that require repeated forward evaluations. Consequently, MD is typically reserved for mechanistic investigations, force-field validation, or targeted refinement rather than routine structural inference. This limitation has motivated the development of alternative dye representations that deliberately balance molecular detail against computational efficiency while preserving predictive capability (**Fig. 3**).

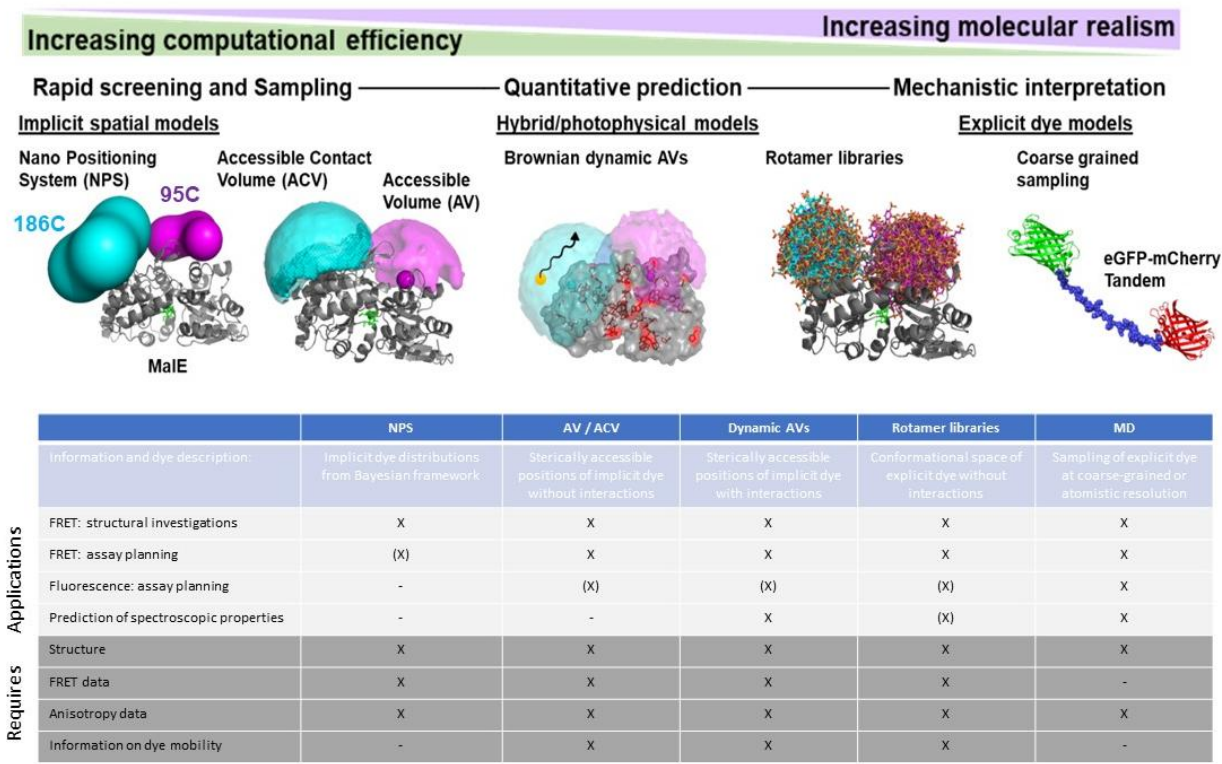

**Figure 3 | Hierarchy of fluorophore forward models in integrative fluorescence modeling.** Dye models translate molecular representations into predicted fluorescence observables linking structural hypotheses to measurements such as FRET efficiencies, fluorescence anisotropies, and quenching rates. Increasing structural detail improves physical consistency, but introduces additional parameters and sampling demands, emphasizing a central principle of integrative structural biology: forward-model complexity should match the information content of the fluorescence experiment to achieve predictive, yet well-constrained structural interpretations. Legend: X: yes, (X) limited, - no.

Geometric modeling strategies like the **Nano Positioning System (NPS)** [37] treat fluorophores as point emitters (satellites) located at unknown, but fixed positions relative to the macromolecule (**Fig. 3**). Using trilateration of multiple distance restraints, NPS localizes dyes and structural features of the host molecule from fluorescence measurements without explicitly modeling linker conformations. The method combines FRET [38] and fluorescence anisotropy [39] in a Bayesian framework to infer spatial probability distributions for dye positions rather than physical occupancy volumes. The core assumption is that dyes occupy quasi-static locations; the inferred distributions therefore reflect positional uncertainty rather than dynamic sampling. NPS is consequently best suited for rigid labeling architectures, i.e., interacting fluorophores with very short linkers, where orientational constraints can be approximated explicitly. Algorithmic refinements have substantially reduced computational cost, yet incorporating dense FRET networks and anisotropy restraints still shifts runtimes from hours to days or longer [39]. Conceptually, NPS is a powerful Bayesian localization approach, but its applicability is restricted: it is not readily transferable to flexible organic fluorophores and is only weakly integrated with broader structural modeling workflows. As such, it is best viewed as a specialized geometric strategy rather than a general fluorophore forward model.

**Accessible-volume (AV)** models represent the most widely-used fluorophore forward models (**Fig. 3**) because they enable rapid and near-exhaustive sampling of the sterically allowed conformational space explored by a tethered dye [40]. AV models approximate the *physical* spatial distribution of the dye a priori and subsequently use this distribution as the forward model. The fluorophore is treated implicitly through an empirical linker length combined with geometric exclusion relative to the macromolecular surface. This allows dye positions to be estimated even when the detailed dye–linker structure is unknown. Accessible contact volume (ACV) models extend this concept by partitioning the AV into mobile and surface-associated fractions, typically guided by fluorescence anisotropy measurements that report on restricted rotation or transient surface interactions (**Fig. 3**). A central advantage of AV approaches is their computational efficiency which makes them attractive for large-scale sampling, Bayesian inference, and integrative workflows in which the forward model must be evaluated repeatedly.

The predictive accuracy of AV/ACV models depends on the parameterization and availability of calibration data. When the assumptions underlying ACV, particularly the extent of surface interactions, are uncertain, this uncertainty can be propagated during the modeling step so that structural conclusions reflect the limitations of the dye description [16]. The orientation factor is therefore typically approximated by the isotropic average, an assumption that is generally valid for small fluorophores undergoing rapid rotational averaging, but may not hold for bulky or strongly interacting dyes. Consequently, AV models are most applicable to small organic fluorophores with substantial rotational freedom. Accurate steric representation likewise benefits from calibration of parameters such as linker length and excluded volume. When experimentally parameterized.

In summary, due to their numerical efficiency the computation of an AV for a given structure requires less than a millisecond and the approach has found widespread applications for many biomolecular systems and notably investigate intrinsically disordered proteins in combination with NMR [19] or as FRET-guided RNA structures modelling [41]. AV-based models have also shown good agreement with measurements and proven reliable in round-robin benchmarks on DNA [26], RNA [41], and, proteins [27]. Importantly, their applicability can often be evaluated experimentally, for example through fluorescence anisotropy, to identify regimes in which restricted dye motion challenges model assumptions.

**Hybrid strategies** address a central limitation of purely geometric AV/ACV models: their inability to capture transient effects arising from dye motion such as dynamic quenching (**Fig. 3**) or other interactions as discussed below. Many fluorescence analyses implicitly assume stationary dye distributions over the fluorescence lifetime and effectively complete rotational averaging. Molecular-dynamics simulations, however, reveal substantial dye displacements on nanosecond timescales [36]. Remarkably, static models often reproduce accurate mean distances, a seeming paradox that underscores the importance of explicitly incorporating dynamics. Hybrid approaches resolve this by augmenting geometric descriptions with stochastic motion [42]. Brownian-dynamics frameworks, for example, simulate linker flexibility and dye diffusion to generate time-averaged positional and orientational distributions while naturally enabling environmental effects such as proximity-dependent quenching. These extend AV-methods occupy an intermediate regime that retains computational tractability yet captures dynamic behavior beyond static volumes, allowing FRET, diffusion, and quenching to be treated within a single physically consistent framework (**Fig. 3**). However, the increased simulation time reduces computational efficiency (from milliseconds for geometric models to seconds) restricting hybrid approaches to experiment design and screening rather than large-scale sampling or Bayesian inference workflows.

**Rotamer**-based approaches seek for even more explicit representations (**Fig. 3**). They describe fluorophores through discrete conformational states rather than implicit spatial envelopes. Typically, conformations of fluorophores and their linkers are selected from a rotamer-library (RL). RL-based approaches were originally developed for electron paramagnetic resonance (EPR) to describe sterically feasible conformations at the labeling site with short spin labels [43]. The RL framework was later adapted to fluorescence applications, where it enabled improved treatment of dye geometry and orientation compared with volume-based models. Because the number of accessible states grows combinatorially with linker length, RL are particularly effective for short to intermediate linkers, for which near-exhaustive sampling remains tractable [44]. Their discrete nature also makes them well suited for forward calculations of distance and orientation distributions while maintaining moderate computational cost. RL are typically atomistic and thus seamlessly integrate into atomistic simulations [45]. A central limitation of rotamer-library approaches is their sampling cost. As a result, rotamer models occupy an intermediate position in the realism–efficiency trade-off: the feature substantially more physical than geometric volume models, are yet markedly cheaper than fully explicit dye simulations, making them especially well-suited for structure selection and ensemble refinement [45].

Beyond geometric AV and RL approaches, **coarse-grained models** provide an intermediate level of representation in which fluorophores and linkers are described by reduced explicit geometries with less detail than full MD (**Fig. 3**). These models retain key physical features, including steric constraints with the host macromolecule, dye-dye exclusion, and orientational preferences, while remaining cheaper than full atomistic simulations. They are particularly useful for sterically demanding labels such as fluorescent proteins (FPs), and restricted organic dyes. The conformational space of the labels can be sampled with the biomolecular system or in a decoupled manner using precomputed ensembles. Simulation-based strategies typically rely on Monte Carlo (MC) or simplified dynamical schemes, which efficiently explore dye conformational landscapes when supported by careful parameterization. Coarse-grained Go models [46], therefore offer a practical compromise between molecular realism and computational tractability. Importantly, the MC sampling does not correspond to a physical timescale. For many bulky fluorophores, particularly FPs, rotational motions are slow relative to the fluorescence lifetime. As a result, orientational effects are better represented by static distributions of the orientation factor rather than by assuming dynamic averaging. This places coarse-grained approaches in a regime well suited for modeling systems where $\kappa^2$ deviations from the isotropic limit may become structurally informative. Consequently, coarse-grained simulations are best viewed as an expensive but informative preprocessing step that constrains the accessible model space rather than as a component of high-throughput scoring.

## Dye models beyond FRET and structural modelling

Dye models and accessible fluorescence data are perquisites for rigorously validated FRET-based structures, yet integrative models use distinct validation criteria from those applied to X-ray crystallography. This largely precluded their broader deposition in the Protein Data Bank [47]. Consequently, the hammerhead ribozyme (PDB ID: 1RMN) is, to the best of our knowledge, the only structure in the PDB primarily derived from FRET restraints [48]. The establishment of PDB-IHM fills this gap by addressing the specific validation and representation requirements of integrative structures [49] providing a formal link between structure and experiment (**Fig. 1/2**) [47].

Dye models were primarily developed for FRET-based structural modelling (**Fig. 3**). Yet, FRET is mostly applied in biochemical assays, imaging and for the study of dynamic processes in biomolecules. All these directions will benefit from accurate dye models. Emerging directions include dynamic accessible volumes, time-resolved rotamer libraries, and hybrid frameworks that link dye simulations on a biomolecule directly to fluorescence observables. Here, a central frontier are dye model improvements for the design of fluorescence assays, which translate desired information, such as the conformational state of a protein, into a spectroscopic observable, i.e., change of solvent-induced quenching (**Fig. 4a**).

Future systematic modelling of fluorescence assays in biology, biochemistry, biophysics, biomedicine and imaging has to consider the treatment of dye orientations, physio-chemical properties and dye interactions with (*i*) the solvent, (*ii*) another dye or quencher and (*iii*) other biomolecules (**Fig. 4a**). In many assays, attachment of organic dyes or fluorescent proteins to a biomolecule is indispensable, therefore labelling sites must be selected that do not interfere with structure and function, e.g., folding, biomolecular interaction or ligand binding (**Fig. 4a**). To support this essential step, we recently introduced the Labelizer (https://labelizer.org/), which ranks proteins residues using a naïve Bayes classifier [28,29], leveraging solvent exposure, conservation, secondary structure and amino acid nature, to rank protein residues in a combined label score. A distribution of label scores is illustrated on the structure of MalE (**Fig. 4b**).

The current Labelizer approach does not consider the nature of the dye, which is, however, a requirement for designing fluorescence assays beyond FRET (**Fig. 4a**). In this context, AV and ACV provide an insufficient compromise between physical realism and computational efficiency since they are based on steric exclusion only. These dye models do not capture dye-interactions, e.g., as required for the establishment of a dye self-quenching assay [50,51], or predict changes in dye-solvent [52] or dye-biomolecule interactions [53] that arise as a consequence of conformational change (**Fig. 4a**). While some coarse-grained models (**Fig. 3**) are suitable to describe orientational effects [54,55], e.g., in homo-FRET [56], they cannot account for minute changes caused by changes in dye properties (charge, spectrum, redox properties, **Fig. 4a**) or proximity of quencher molecules such as tryptophans (**Fig. 4a**). Inherently, hybrid AV models are capable to account for quantum yield changes of dyes by amino-acid quenching [42]. Importantly, aspects such as the solvent environment are not yet considered, but represent flexible (continuous) parameters to optimize modelling of spectroscopic data for assay design.

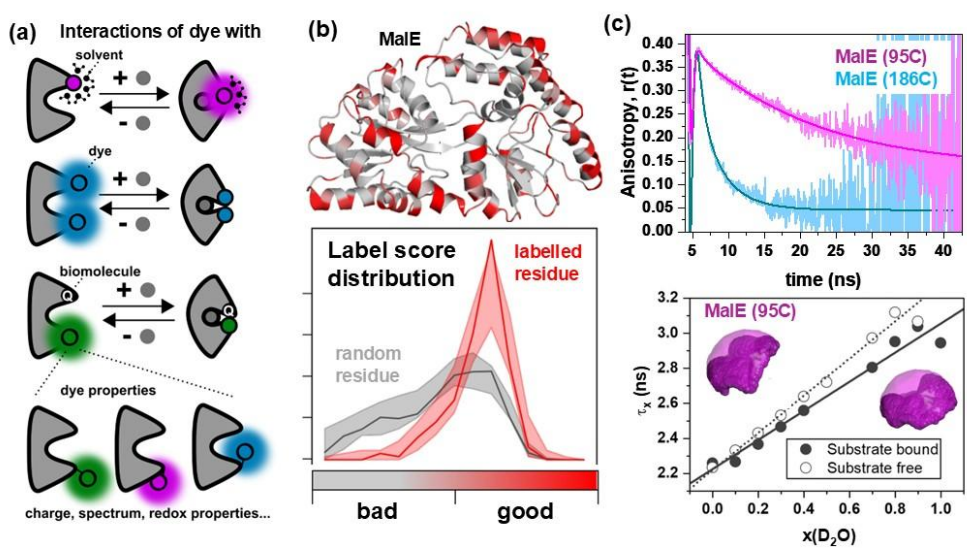

**Figure 4 | Suggested parameters for integration into hybrid AV models of dyes for systematic planning of fluorescence assays.** (a) Selection of suitable residues on biomolecules considering distinct dye interactions and properties. (b) Selection of suitable residues for dye attachment on proteins using MalE of E. coli (pdb:1OIB) as an example for suitable (red) and non-suitable (grey) residues. (c) Time-resolved fluorescence anisotropy decays (top) for labeling positions 95C and 186C show site specific rotational behavior. The dependence of the average fluorescence lifetime $\tau_x$ on the $D_2O$ fraction shows a clear maltose induced difference in the solvent exposure (bottom). Holo MalE exhibits a stronger increase in $\tau_x$ with $D_2O$. Accessible volume (AV) solvent exposure explain lifetime effects: Open MalE (PDB-ID: 1FQA) yields a surface volume of ~2.56 nm³ and an interior volume of ~6.09 nm³ (surface/volume ratio 0.42). Maltose-bound MalE (PDB-ID: ANF1) shows a similar surface volume (~2.59 nm³) but a larger interior volume (~8.90 nm³), resulting in a lower solvent exposure ratio (0.29) and weaker $D_2O$ sensitivity.

We illustrate the necessity for advanced dye models using unpublished preliminary data on position- and state-dependent effects in the maltose binding protein related to two well-described variants: MalE-D95C [57] and MalE-A186C [28]. The time-resolved fluorescence anisotropy decays of ATTO655 on both labeling positions show site specific rotational behavior, as expected for a restricted (D95C) and free (A186C) dye position (**Fig. 4c**). Time-resolved anisotropy measurements indicate comparable dye mobility, however, in the two conformational states of both protein variants (data not shown), excluding substantial changes in rotational dynamics. These spectroscopic parameters thus fail to explain the ability of D95C to serve as a conformation-specific position [57]. Also, the attempt to analyze classical AVs is unsuccessful since the AV surface volumes in both conformational states are equal. Thus, a central question for assay design becomes which modelling parameter can predict the ability of residues to sense conformational change upon maltose binding via spectroscopic indicators.

Importantly, we noted a significant difference in the response of the fluorescence lifetimes of both conformational states when water is substituted by $D_2O$, which removes water-induced singlet-quenching (**Fig. 4a**). The maltose-free state displays a stronger lifetime increase with increasing molar fraction, $x(D_2O)$, than the maltose-bound state. Because $D_2O$ sensitivity scales with solvent-mediated quenching [52], the larger effect in the unbound apo-state is related with an increased solvent exposure. To account for this, we analyzed the surface-to-volume ratio (SVR) of the AVs: MalE apo (PDB-ID: 1FQA) has a surface volume of ~2.56 nm³ and an interior volume of ~6.09 nm³ (SVR 0.42), while MalE holo (PDB-ID: ANF1) shows a similar surface volume (~2.59 nm³) but a larger interior volume (~8.90 nm³), resulting in a lower SVR 0.29. This can explain the weaker $D_2O$ sensitivity (**Fig. 4c**, bottom), since ACV analysis separates the total number of accessible conformational points from the solvent-exposed subset and enables calculation of the SVR. Comparison of the maltose-free and maltose-bound states shows that the maltose-free protein exhibits a higher SVR, indicating that a larger fraction of dye conformations is solvent exposed. Upon maltose binding, this ratio decreases, consistent with a more internally confined accessible volume.

**Acknowledgements**

This work was generously supported by start-up funding of TU Dortmund University and Deutsche Forschungsgemeinschaft (DFG-CO879-6-1, project Linker to T.C.).

**Author contributions**

T.O.P. and T.C. discussed the content of the article and prepared figures. T.O.P. wrote the initial draft of the paper, which was finalized with T.C. D.M. conducted experiments and provided data. T.C. acquired funding. All authors discussed, edited and contributed to the final version of the paper.

**Competing Interest Statement**

T.C. is a scientific co-founder and share-holder of FluoBrick Solutions GmbH a company that distributes fluorescence microscopy and spectroscopy instruments.